\documentstyle[preprint,aps,epsfig]{revtex}

\begin{document}



\title{Apparent hysteresis in a driven system with self-organized drag}

\author{Mikko Haataja$^{1,}\footnote{Current address:  Department of Materials
Science and Engineering, McMaster University, 1280 Main Street
West, Hamilton, Ontario, Canada L8S 4L7.}$, David J.
Srolovitz$^{1}$, and Ioannis G. Kevrekidis$^{2}$}
\address{$^1$ Princeton Materials Institute and Department of Mechanical
and Aerospace Engineering, Princeton University, Princeton, New
Jersey 08544, USA.}
\address{$^2$ Department of Chemical Engineering, PACM and
Mathematics, Princeton University, Princeton, New Jersey 08544,
USA.}

\date{\today}

\maketitle

\vskip0.5cm

\begin{abstract}
Interaction between extended defects and impurities lies at the
heart of many physical phenomena in materials science. Here we
revisit the ubiquitous problem of the driven motion of an extended
defect in a field of mobile impurities, which self-organize to
cause drag on the defect. Under a wide range of external
conditions (e.g. drive), the defect undergoes a transition from
slow to fast motion. This transition is commonly hysteretic: the
defect either moves slow or fast, depending on the initial
condition.  We explore such hysteresis via a kinetic Monte Carlo
spin simulation combined with computational coarse-graining.
Obtaining bifurcation diagrams (stable and unstable branches), we
map behavior regimes in parameter space. Estimating fast-slow
switching times, we determine whether a simulation or experiment
will exhibit hysteresis depending on observation conditions.  We
believe our approach is applicable to quantifying hysteresis in a
wide range of physical contexts.
\end{abstract}

\vskip0.5cm

\narrowtext

Driving an extended defect or domain wall in a system that
contains stationary or mobile impurities with which the domain
wall interacts lies at the heart of many physical phenomena,
including the motion of grain boundaries in polycrystalline
materials \cite{cahn62}, the motion of vortex lines in dirty
superconductors \cite{blatter94,chauve00}, driven charge-density
waves \cite{gruner88}, the motion of dislocations in impure metals
undergoing plastic deformation (i.e., Portevin-LeChatelier effect
\cite{hahner02}), ferroelectric domain wall dynamics
\cite{tybell02}, and stick-slip phenomena in tribology
\cite{cule96}. The ability to quantitatively model such systems
holds the key to designing a wide range of devices and optimizing
their operating conditions. While understanding the transition
between the slow and fast kinetic regimes is vital, the dynamics
become particularly difficult to characterize in the neighborhood
of such a transition (neutral stability).

When equations that describe the macroscopic behavior of a system
are known analytically, an established set of
mathematical/computational tools is available for analyzing the
dynamics (e.g., locating and characterizing transitions). Many
``real world'' applications are too complex to characterize in
terms of simple equations. Microscopic simulations are becoming an
increasingly common approach to analyze situations where such
macroscopic models are not available analytically. This is the
case for driving a domain wall through a field of mobile
impurities: although analytical models exist, they do not
accurately and/or fully predict the range of observed behavior. In
such systems, there is a ``specter'' of hysteresis in the
transition from slow to fast motion -- sometimes it is seen
\cite{maroudas91} and sometimes not \cite{mendelev01}. Classical
theories of such phenomena  \cite{cahn62} are usually
one-dimensional and show that the domain wall velocity vs. driving
force curve can be multi-valued, i.e., suggest the existence of
hysteresis. In this brief report, we examine this phenomenology --
try to understand its roots, and argue that it is both the physics
{\it and} the observer that determines whether hysteresis will
appear in experiments or simulations.

We illustrate the issues arising in considering hysteresis within
the context of the classical solute drag effect, in which
diffusing impurities interact with a migrating domain wall and
impede its motion. In particular, we employ a variant of the Ising
model in which the domain wall moves under the influence of an
external field, and interstitial impurities diffuse and are
attracted to the domain wall \cite{mendelev01}. The energy of the model
system is ${\cal{H}}/k_BT = -J/2 \sum s_i s_j-h
\sum s_i -E_0/4 \sum \epsilon_\alpha |\sum s_j|$, where $s_i=1$ in
one domain and $-1$ in the other, $\epsilon_\alpha$ is zero or one
depending on whether an impurity occupies interstitial site
$\alpha$ or not, $E_0>0$ implies that the impurities are attracted
to the interface, and $J>0$ and $h$ scale the domain wall energy
and the driving force, respectively. The first term represents the
total energy of domain walls in the pure system, the second term
accounts for the external drive, and the last term describes the
interaction between the impurities and the domain wall. The
parameters chosen for the present study were $J=2.0$ and $E_0 =
[4.0,7.0]$ and the bulk impurity concentration was set to 1{\%}
($c_{imp}=0.01$).

Initially, the 2-d  computational domain contains a single,
straight domain wall and randomly placed interstitial impurities.
The system evolves by randomly choosing spins to flip or
impurities to exchange with neighboring empty sites, with
probability proportional to their relative density and the
impurity diffusivity $m$, as described by Mendelev, et
al.\cite{mendelev01}. An attempted change is accepted if it lowers
the energy of the system ($\Delta H \le 0$) or, if $\Delta H >0$,
it is accepted provided that e$^{-\Delta H/k_BT} <R$, where $R$ is
a random number uniformly distributed between 0 and 1. The
simulation clock advances by one unit when all of the spins have
been sampled once, on average. When the domain wall gets close to
one end of the simulation cell, more sites are added to that end
and removed from the opposite end, thereby making the length of
the system $L$ effectively infinite \cite{mendelev01}. The results
presented below are for a relatively small simulation cell width
($W=32$). This was done to prevent excessive domain wall
roughening, which tends to smear the slow-fast domain wall
transition. The effect of system size is very important for the
observation of hysteresis, but will be the subject of a future
report.

The domain wall velocity results from a balance between the
external field, which controls the propagation velocity in the
pure system, and the diffusing impurities, which locally impede
domain wall migration. In the limit of large drive and weak domain
wall-impurity interactions, the impurities have little influence
on the domain wall motion. Conversely, decreasing the drive and
increasing the attraction of the impurities to the domain wall
$E_0$ leads to very slow domain wall dynamics, due to the strong
segregation of the impurities onto the domain wall. However, as
will be demonstrated below, the drive, impurity-domain wall
interaction, impurity mobility and the average impurity
concentration may conspire to give rise to an intermediate regime:
a regime where the domain wall is captured by the impurities, then
escapes and travels nearly unimpeded  until it is recaptured. This
is the ``jerky'' (stick-slip-like) domain wall motion regime,
where hysteresis may be observed.

Figure~\ref{fig:varh} shows the mean domain wall position $R$ vs.
time $t$ for $h=0.12$ and several values of the domain
wall/impurity interaction strength $E_0$.
These curves become qualitatively different as $E_0$ is changed.
In particular, for $E_0=4.0$, the domain wall propagates smoothly,
while increasing $E_0$ from $E_0=5.5$ to $E_0=6.5$ produces
``jerky'' motion, in which the domain wall slows dramatically for
long time intervals before breaking free again. Examination of
Fig.~\ref{fig:varh} also shows that intervals of slow motion are
strongly correlated with the presence of a large number of
impurities on the domain wall, $N$, and {\it vice versa}. Transitions
in domain wall speed are strongly correlated with a jump in the number
of impurities on the domain wall.

It is instructive to consider how the changes in $R(t)$ are
reflected in the distribution of the number of impurities on the
domain wall $P(N)$, shown in Fig.~\ref{fig:pdist}, obtained from
long-time simulations at several $E_0$. For relatively weak domain
wall/impurity interactions ($E_0=5.25$), $P(N)$ is unimodal -- the
system is (almost always) in the fast regime. As the interaction
strength is increased, $P(N)$ becomes bimodal as the system
transits between the slow and fast regimes (i.e., jerky motion).
Finally, for sufficiently strong interactions, $P(N)$ is again
unimodal, with the system effectively trapped in the slow state.

In the ``jerky'' regime, observations of the system can be
characterized by two distinct time scales, namely $t_S$,  the
characteristic time  that the domain wall motion remains slow
(effectively trapped, many impurities on the domain wall) before
it breaks free; and $t_F$, the characteristic time that the domain
wall motion is fast (few impurities on the domain wall) before
again being captured by the impurities. If our observation time
$t_O$ is smaller than both these characteristic times, the system
will {\it appear to us} in a typical simulation to be either only
fast or only slow, depending on the initial condition (suggesting
hysteresis). On the other hand, if $t_O$ is much greater than
these times, the domain wall appears to move at a constant speed
that is an appropriately weighted average of the fast and slow
speeds. The motion appears to be ``jerky'' only if $t_O$ is in the
range of the two characteristic times.

In what follows, we focus on the variable $N(t)$ and demonstrate
that a quantitative description of the system can be developed
based upon its dynamics. Assuming that $N(t)$ is an effective
``slow variable'' for the system, standard arguments based on the
Mori-Zwanzig projection operator formalism (see, e.g., Ref.
\cite{forster}) lead to a stochastic description of the system in
terms of a Langevin equation for $N(t)$. Alternatively, we may
consider a one-dimensional Fokker-Planck (F-P) equation,
describing the statistics of the stochastic motion of a
``particle'' in a two-well potential; here the ``particle''
corresponds to the instantaneous number of impurities on the
domain wall $N(t)$, and the effects of all other degrees of
freedom are incorporated into the effective potential. The
effective F-P equation for the corresponding probability
distribution $P(N,t)$ is \cite{kubo}
\begin{equation} \label{eq:fp}
\frac{\partial P(N,t)}{\partial t} = \left[
-\frac{\partial}{\partial N} V(N) + \frac{\partial^2D(N)}{\partial
N^2} \right]P(N,t),
\end{equation}
where $V(N) \equiv \lim_{\Delta t \rightarrow 0} \langle \Delta N
\rangle/ \Delta t$, $2D(N) \equiv \lim_{\Delta t \rightarrow 0}
\langle [\Delta N]^2 \rangle/ \Delta t$, and $\Delta N \equiv
N(t+\Delta t)-N(t)$. At stationarity, $P(N) \propto
\exp(-\Phi(N)/k_BT)$, where $\Phi(N)$ denotes the effective
potential and is determined from $V$ and $D$ as
\begin{equation} \label{eq:g}
\frac{\Phi(N)}{k_BT} = const.  -  \int_0^N dN' \frac{V(N')}{D(N')} +
\ln D(N).
\end{equation}

Note that, in the limit $D=const.$, Eq.~(\ref{eq:fp}) becomes
equivalent to the Langevin equation $\frac{\partial N}{\partial
t}= -\frac{D}{k_BT} \frac{d \Phi}{dN} + \eta$, where $\eta$ is a
random, stochastic variable that is Gaussian distributed with mean
$\langle \eta \rangle = 0$ and variance $\langle \eta(t) \eta(s)
\rangle = 2 D \delta(t-s)$.  Most importantly, our MC simulations
can be employed to extract both $V(N)$ and $D(N)$ ``on demand'' as
follows: choose a value of $N_0$, locate several instances when it
appears in a long time simulation, tabulate the subsequent values
of $N$ within a fixed time interval $\Delta t$ (here $\Delta
t=600$), and then average over these segments and estimate the
rate of change in the mean ($V$) and the variance ($2D$) for the
number of impurities on the domain wall. This is repeated for a
grid of  $N_0$ values sufficient to numerically evaluate the
integral in Eq. (\ref{eq:g}) to the desired accuracy. Similar
methods for estimating $V$ and $2D$ from a stochastic time series
has been presented in Refs. \cite{gradisek00,hummerk}.

The result of this analysis is shown in Fig.~\ref{fig:geff}, where
we plot $\Phi(N)/k_BT$ for $E_0=6.0$ along with the potential
obtained by directly constructing the probability distribution
$P(N)$ from the time series and employing $P(N)\sim
\exp(\Phi/k_BT)$. Indeed, $\Phi(N)/k_BT$ has a double-well
structure, with minima at $N_{F} \approx 4$ and $N_{S} \approx
24$, and an (unstable) saddle transition point at $N_U=11$.
Furthermore, the effective potential extracted from the F-P
analysis shows good agreement with that obtained directly from the
time series for $N>6$. The discrepancy observed for small $N$ can
be rationalized by arguing that $\Phi(N)$ is no longer effectively
one-dimensional. In this regime, additional degrees of freedom,
such as the domain wall shape, are no longer slaved to the single
variable $N(t)$.

The form of the effective potential $\Phi(N)$ suggests that there
is a single effective barrier or saddle that must be overcome for
the system to move from the slow (pinned) state to the fast one
(or {\it vice versa}).  The waiting time between transitions
between states depends on the shape of $\Phi(N)$ and the kinetic
coefficient $D(N)$. Following Kramers \cite{hanggi90}, we can
estimate this time $\tau$, as
\begin{equation} \label{eq:Kramer}
\tau \approx   \frac{2\pi k_B T}{ \bar{D} \sqrt{\Phi^{''}
(N_{min}) |\Phi^{''}(N_{saddle})|}}e^{\Delta \Phi/k_B T},
\end{equation}
where $2\bar{D}=D(N_{min})+D(N_{saddle})$. For the data
represented in Fig.~\ref{fig:geff} corresponding to $E_0=6.0$, the
average time needed for a transition from fast to slow and
{\it{vice versa}} is $\tau \approx 4 \times 10^4$ and $\tau
\approx 4 \times 10^5$, respectively. These are each within a
factor of $3$ of the waiting times estimated from a long MC
simulation.

Armed with $\Phi(N)$ for several values of $E_0$, we now construct
the domain wall speed vs. impurity interaction strength, $E_0$,
bifurcation diagram, as follows.  We determine the average domain
wall speed  as a function of $N$, $\bar{V}=\bar{V}(N)$, for each
$E_0$ from our numerical simulations. Then, we determine the
velocity on the slow branch as $V_S=\bar{V}(N=N_{S})$, on the fast
branch as $V_F =\bar{V}(N=N_{F})$, and on the unstable branch as
$V_U = \bar{V}(N=N_{U})$, where $N_{S}$, $N_{F}$, and $N_{U})$
from plots such as those in Figs.~\ref{fig:pdist} and
\ref{fig:geff} for each value of $E_0$. The results are shown in
Fig.~\ref{fig:bif}, where we plot the stationary domain wall
velocity vs. interaction strength bifurcation diagram, including
multiple stable branches and the unstable branch.  Alternatively,
we can search for the stationary values of $N$ for which $dN/dt=0$
using standard root-finding methods. $dN/dt$ can be estimated as
$\frac{N(t=0)-N(t=t^\prime)}{t^\prime}$ by performing a series of
short simulations initialized with $N(t=0)$ impurities on the
boundary for a time $t^\prime$ \cite{hummerk}.

To summarize, we obtain both the stable and unstable branches of
the bifurcation diagram, as well as an estimate of the
characteristic time $\tau$ over which the domain wall will switch
from fast to slow or {\it vice versa}.  Therefore, hysteresis can
be expected for observation times $t_O \ll \tau$, and the
bifurcation diagram becomes single-valued everywhere for
observation times $t_O \gg \tau$.  In the latter limit, the
observed domain wall velocity will be $\bar{V}(E_0)=\int_0^\infty dN
P(N;E_0)\bar{V}(N;E_0)$, as shown  in Fig.~\ref{fig:bif}.
Increasing either the impurity diffusivity or the heat of
segregation (or both) leads to more effective pinning, and thus
increasingly jerky domain wall dynamics.  In Fig.~\ref{fig:bif} we
show a morphological diagram for the propagation mode of the
domain wall as a function of $m$ and $E_0$ for $t_O=10^4$.

Combining a coarse-grained computational approach with an {\it
effective} Langevin/Fokker-Planck description for the stochastic
dynamics of the impurity density along the domain wall provides  a
useful and computationally efficient method for determining the
apparent stationary states (both stable and unstable) of the
system, as well as for computing the transition rates between the
stable states.  The transition rates, together with the apparent
stationary states, provide a clear picture of the dynamics of the
coupled domain wall-impurity system and conditions for determining
when hysteretic behavior will be observed.  We believe that this
approach can be used to quantifying the emergence of hysteresis in
a wide variety of physical systems. Whether hysteresis will be
observed in a particular experiment or simulation depends, of
course, on the physics.  However, it does not only depend on the
physics; it depends on the time scale of the observation and on
the (length, ensemble) size of the system.  In this sense,
{\it{hysteresis is in the eye of the beholder.}}

\newpage

\begin{figure}[!]
\caption{ (a) Position of the domain wall as a function of time
for several values of the heat of segregation, $E_0$.  (b) The
number of impurities $N(t)$ on the domain wall, corresponding to
the $E_0=6.0$ simulation in (a). (c,d) Configurations of the
domain wall and impurity positions at the times marked ``X'' and
``O''  in (a), respectively.} \label{fig:varh}
\end{figure}

\begin{figure}[!]
\caption{Steady-state probability distributions $P(N)$ as a
function of $E_0$, determined from long MC runs.}
\label{fig:pdist}
\end{figure}

\begin{figure}[!]
\caption{The effective potential $\Phi(N)$ from Eq. (2) and
directly from the long-time simulation data of
Fig.~\ref{fig:pdist} with $E_0=6.0$.}
\label{fig:geff}
\end{figure}

\begin{figure}
\caption{(a) The effective domain wall velocity versus $E_0$
bifurcation diagram at $m=0.25$. The solid and dashed lines are
guides to the eye. (b) A diagram of the different regimes of
behavior as a function of impurity mobility $m$ and the
impurity-domain wall interaction strength $E_0$; blue denotes
smooth, gray jerky, and red effectively pinned propagation.
Symbols are from measured data with ``F'' corresponding to fast,
``J'' to jerky, and ``S'' to slow (pinned) behavior, respectively.}
\label{fig:bif}
\end{figure}


\end{document}